\documentclass[%
superscriptaddress,
 amsmath,amssymb,
 aps,
showkeys
]{revtex4-2}

\usepackage{graphicx}% Include figure files
\usepackage{dcolumn}% Align table columns on decimal point
\usepackage{bm}% bold math
\usepackage{xspace}
\usepackage{xcolor}
\usepackage{ulem}
\usepackage{definitions}
\usepackage[mathlines]{lineno}% Enable numbering of text and display math
\begin{document}

\preprint{APS/123-QED}

%\title{Manuscript Title:\\with Forced Linebreak}% Force line breaks with \\

\title{Can Transverse Mass Scaling Shed Light on the Event-Activity Dependence of $\ensuremath{\mit{\Upsilon}}$ Mesons Production at LHC?}%

\author{Iakov Aizenberg}
\affiliation{Department of Physics and Astrophysics, Weizmann Institute of Science, Rehovot, 761001, Israel}
\author{Zvi Citron}%
\affiliation{Department of Physics, Ben Gurion University of the Negev, Beer Sheva, 8410501, Israel}%
\author{Alexander Milov}
\email{alexander.milov@weizmann.ac.il}
\affiliation{Department of Physics and Astrophysics, Weizmann Institute of Science, Rehovot, 761001, Israel}

\date{\today}

\begin{abstract}
Measurements by the CMS experiment~\cite{CMS:2013jsu,CMS:2020fae} reveal a deficit of charged particle tracks in events with higher $\ensuremath{\mit{\Upsilon}(\text{nS})}$ states. This observation is suggested to be a manifestation of the excited bottomonia suppression in $pp$ interactions. Transverse mass ($m_{\mathrm{T}}$) scaling can be implied to check this assumption in an independent way.  The scaling has been observed for a wide range of particle species in proton-proton collisions at various energies from the SPS to RHIC and the LHC. The observed scaling is known to be different for baryons and mesons, and this work presents a comprehensive study of the  $m_{\mathrm{T}}$-scaling of mesons at LHC energies with a focus on heavier mesons. The study demonstrates patterns in the scaling properties of mesons, which are related to the particle quark content. 
In particular, light species and ground-state quarkonia obey the same scaling, whereas open-flavor particles deviate from it because their spectra are significantly harder. The magnitude of deviation depends on the flavor of the heaviest quark in the meson. By extending the $m_{\mathrm{T}}$-scaling assumption to the excited bottomonia states, it is observed that the measured cross sections of $\Upsilon$(2S) and $\Upsilon$(3S) are reduced by factors of 1.6 and 2.4 compared to the expectation from the scaling.  This observation is consistent with recently observed differences between the event-activity dependence of different $\Upsilon$(nS) meson states. 
\end{abstract}


\maketitle

\section{\label{sec:intro} Introduction}

Based on a statistical thermodynamical approach, it was suggested~\cite{Hagedorn:346206} that hadron production in proton-proton (\pp) collisions scales with the transverse mass of the produced particles. Transverse mass is defined as $\mT=\sqrt{\pT^{2}+m_{0}^{2}}$, where \pT is the momentum of the hadron in the plane orthogonal to the collision axis, and $m_{0}$ is its rest mass. This scaling was demonstrated experimentally in measurements at the ISR experiment~\cite{Alper:1973xrf,ALPER197375}. Although no longer thought to represent a fundamental hadronic temperature as originally proposed, it has since been used by many experiments and phenomenological studies to understand particle production in \pp and nucleus-nucleus collisions from the SPS to RHIC and the LHC~\cite{Bratkovskaya:1997dh,STAR:2006nmo,STAR:2006nmo,Altenkamper:2017qot}. %In its more generalized form based on Tsallis statistics~\cite{Tsallis:1987eu}, the original exponential form is often replaced with a power law, which better describes spectra at higher collision energies (\sqs).
%The original expression of \mT-scaling used an exponential form~\cite{Hagedorn:346206}, but it may be expressed in a more generalized form based on Tsallis statistics~\cite{Tsallis:1987eu} in which a power law is used providing a better description of particle spectra at higher collision energies (\sqs).
The original expression of \mT-scaling~\cite{Hagedorn:346206} used an exponential form, but it may also be derived based on Tsallis statistics~\cite{Tsallis:1987eu} in which a power law provides a better description of particle spectra at higher collision energies (\sqs).
Various extensions of the transverse-mass scaling incorporate different assumptions and are capable of describing a broad variety of experimental data with impressive precision~\cite{Grigoryan:2017gcg}. In this letter, an assumption of \mT scaling is used to provide a qualitative corollary to measurements of \Upsa mesons made by the CMS experiment in \pp collisions~\cite{CMS:2013jsu,CMS:2020fae}. CMS observed an intriguing correlation between the order of the \Upsn-meson state and the multiplicity of charged hadrons measured in \pp collisions and suggested that this may correspond to a suppression of the excited \Upsn states.
In this analysis, the \mT-scaling assumption is used to define a baseline for excited \Upsa meson production from which suppression could be estimated.

The differential production cross-section can be approximated with the form 
\begin{equation}
\frac{\mathrm{d}^{2}\sigma}{\mathrm{d}y\mathrm{d}\mT} \propto \left(1+\frac{\mT}{nT}\right)^{-n}
\label{eq:mT}
\end{equation}
which is derived from Tsallis statistics~\cite{PHENIX:2010qqf,Grigoryan:2017gcg}. In addition to \mT, the cross-section in the left-hand side of Eq.~\eqref{eq:mT} is written as differential also in rapidity, $y = \frac{1}{2} \ln\left[(E+p_{z}) / (E-p_{z})\right]$, where $E$ is the energy of a particle and $p_{z}$ is the momentum component in the direction of colliding protons. In principle, $y$ factorizes from Eq.~\eqref{eq:mT}, but in practice, a weak dependence of the parameters in the right-hand side of Eq.~\eqref{eq:mT} on $y$ remains in the data~\cite{Grigoryan:2017gcg}. Transverse mass scaling assumes that the exponent $n$ and parameter $T$ are universal for all particles for a given \sqs.  It has been shown in many papers that mesons and baryons do not obey the same scaling (e.g., \cite{Altenkamper:2017qot} and references therein). The analysis presented in this letter considers only mesons. The scaling behavior of heavy mesons is studied in comparison to light quark species and is used to understand the ratios between excited and ground quarkonia states measured in experiments.

\section{\label{sec:data}  Experimental Data}
The experimental data used in this work is drawn from \pp collisions at  $\sqs=7$, 8, and 13~TeV measured by the ALICE~\cite{ALICE:2012wos,ALICE:2016dei,ALICE:2015ial,ALICE:2020ylo,ALICE:2020jsh,ALICE:2019hyb,ALICE:2017olh,ALICE:2014uja,ALICE:2017ryd,ALICE:2015pgg,ALICE:2021dtt,ALICE:2017leg,ALICE:2021ucq}, ATLAS~\cite{ATLAS:2015igt,ATLAS:2011aqv,ATLAS:2014ala,ATLAS:2015zdw,ATLAS:2013cia,ATLAS:2012lmu,ATLAS:2011xhu,ATLAS:2016kwu}, CMS~\cite{CMS:2010nis,CMS:2015lbl,CMS:2011pdu,CMS:2011oft,CMS:2011kew,CMS:2013qur,CMS:2015xqv,CMS:2011rxs,CMS:2017eoq,CMS:2017dju,CMS:2016plw}, and LHCb~\cite{LHCb:2013xam,LHCb:2011zfl,LHCb:2012geo,LHCb:2013vjr,LHCb:2015log,LHCb:2011ijs,LHCb:2019eaj,LHCb:2013itw,LHCb:2015swx,LHCb:2019zaj,LHCb:2015foc,LHCb:2018yzj} experiments. %The data (including their uncertainties) are obtained from the HEP database. The data is processed using the following procedure: 
The data are obtained from the HEP database and processed using the following procedure: 
\begin{itemize}
\item[$\cdot$] Only measurements of mesons are considered.
\item[$\cdot$] Where available, the prompt production component is used. For charmonia states, this excludes feed-down contributions from $B$ mesons but includes the feed-down from decays of heavier charmonium states. These processes may contribute more than 30\% of the measured charmonium cross section~\cite{CMS:2015lbl}, and the analogous contribution to bottomonium states may be even larger ~\cite{LHCb:2014ngh}.
\item[$\cdot$] For measurements which are corrected using multiple meson polarization assumptions, the results corresponding to the zero polarization assumption are selected.
\item[$\cdot$] Measurements reported within a detector's fiducial acceptance are corrected to the total cross-sections using the zero polarization assumption.
\item[$\cdot$] Since particle ratios for quarkonia states considered in this letter are typically measured in dilepton decays channels, the particle ratios also include branching ratios.
\item[$\cdot$] Cross-sections measured to rapidity larger than 3 but reported differentially in rapidity are selected with $|y|<3$ for better consistency with the results measured at mid-rapidity.
\end{itemize}

The collected data contains 88 independent measurements of particle species and 18 measurements of the particle ratios, spanning transverse momentum from 0 to 150~GeV and absolute rapidity from 0 to 4.5. Measurements that contain less than 5 data points or span less than several GeV in transverse momentum are not used in the analysis. Where multiple measurements of the same particle are reported by the same experiment at the same energy, the measurements typically rely on different statistical samples and therefore are considered to be independent. They are combined using a constant scaling factor applied to one of the measurements. This factor is calculated to account for small differences in the properties of the different measurements, for example, the rapidity coverage. The magnitude of the factor is close or equal to unity. Measurements of isospin-partner particles made by the same experiment and measurements of the same particles performed by different experiments are not combined.

Altogether, this analysis uses 72 combined data samples of 18 particle species and their isospin partners with 1509 experimental data points and 15 measurements of particle ratios with 327 data points.

\section{\label{sec:fits} Fits}
Each data sample is fitted to the functional form given by Eq.~\eqref{eq:mT}. 
Point-by-point uncorrelated uncertainties are used in the fits. Point-by-point fully correlated uncertainties (scaling factors) are not considered because the spectra of different particles are re-normalized in the analysis. Other point-by-point correlated uncertainties are shown in the relevant figures but are not input into the fits. 

In addition to the general data handling discussed above, several data exclusions are made in the fitting procedure. Measurements of $\pi^{0}$ and $\pi^{\pm}$ with $\pT<2$~GeV are excluded based on the considerations discussed in Refs.~\cite{Altenkamper:2017qot,ALICE:2020jsh}. 
Low momentum, $\pT<5$~GeV, \Upsa meson measurements are also excluded from the fitting because in the low momentum region, there is a large contribution from $\chi_{b}(\mathrm{mP})\rightarrow\Upsn\gamma$ decays with $m\geq n$. A correction for feed-down decays from heavier mesons ($\chi_{c}$, $\chi_{b}$) is not applied because it cannot be reliably determined from existing data, although it is expected to be similar in magnitude for all \Upsn states~\cite{Andronic:2015wma}. 

\begin{figure}[htb]
\includegraphics[width=0.8\textwidth]{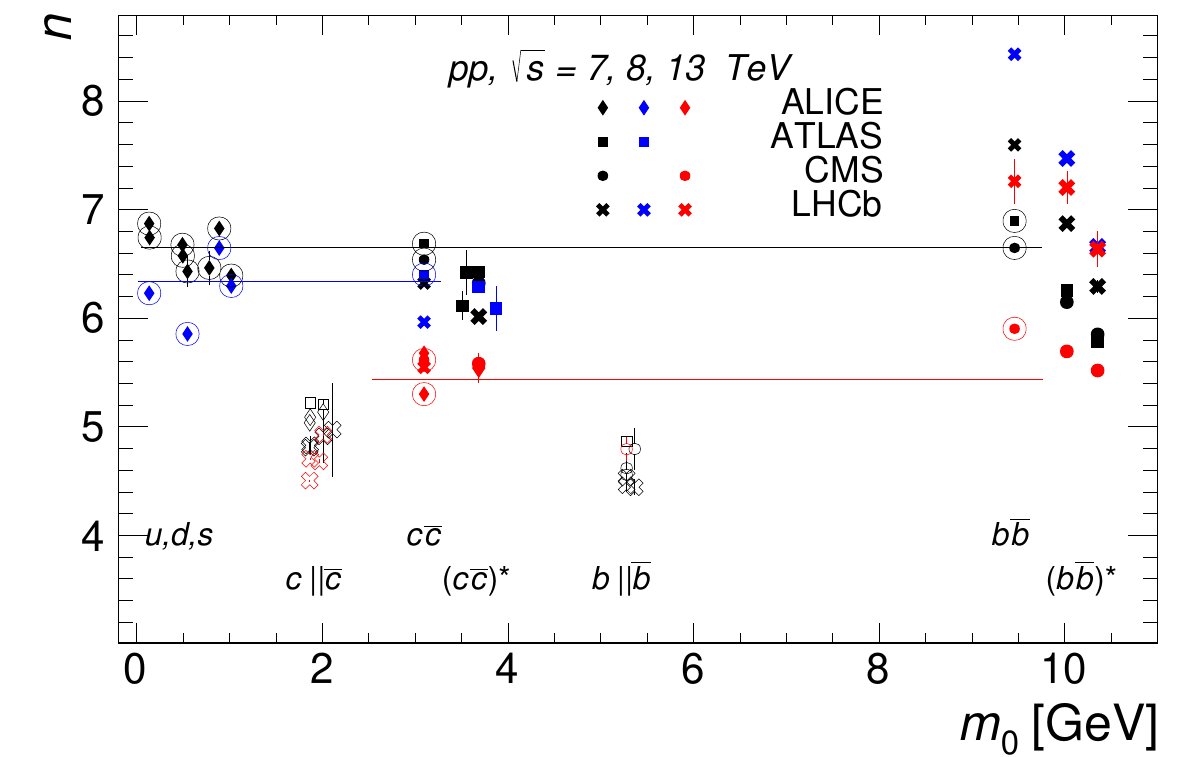}
\caption{\label{fig:individual_fits} Color online. The mass dependence of the \mT-spectra exponent $n$ for different measured species. Symbol shapes denote experiments, open symbols denote open heavy flavor particles, and larger symbols denote excited quarkonia states. Error bars are uncertainties of the fits. Encircled points indicate the results which are used for common fits of the \uds and ground-state \qum mesons.  The results of these fits are shown by horizontal lines. Different colors represent different collision energies.}
\end{figure}

Since the fit parameters $T$ and $n$ are strongly correlated with each other, unconstrained fits produce results that are difficult to interpret in a coherent way. Therefore, the parameter $T$ is fixed to 254~MeV for all analyzed particles. This value is obtained from a simultaneous fit to several selected data sets at $\sqs=7$~GeV and is close to the value of the same parameter used in~\cite{Altenkamper:2017qot}. The exponent, $n$, obtained from the fits is plotted versus the particle rest mass in Figure~\ref{fig:individual_fits}.

There are several patterns visible in the figure. Particles produced in collisions with higher \sqs have harder spectra which corresponds to lower values of $n$, given a fixed $T$. LHCb results for \Upsa are higher at each collision energy because they are measured at higher rapidity $2<|y|<3$. Comparing results at the same energy shows that $n$ increases with rapidity, reflecting the fact that particle spectra become softer at high rapidity~\cite{Grigoryan:2017gcg}. Several other measurements of light mesons that are made at high rapidity also show higher $n$, they are not plotted in Figure~\ref{fig:individual_fits} for plot clarity.

The magnitude of $n$ obtained from the fits shows a strong dependence on the quark content of the particle. Open heavy flavor mesons (\ohf) demonstrate much harder spectra compared to other species. This holds for all energies, and the effect is stronger for open bottom (\obtm) than for open charm (\ochm).

As has been observed in~\cite{Grigoryan:2017gcg}, at LHC and lower energies, light particle species (\uds) and quarkonia states follow \mT-scaling, i.e., their spectral distributions can be fitted to Eq.~\eqref{eq:mT}, or to the Tsallis form~\cite{Tsallis:1987eu}, with the same parameters $T$ and $n$. In a large number of publications, the same statement is derived for narrower selections of mesons, e.g., in~\cite{ALICE:2020jsh,PHENIX:2010qqf,STAR:2006nmo}.

Figure~\ref{fig:individual_fits} shows that the values of $n$ for \uds and ground-state heavy quarkonia (\qum) mesons are similar. However, excited quarkonia states (\equm) have lower values of $n$ than the ground state as seen by the ordering $n(\Uone)>n(\Utwo)>n(\Uthree)$. A similar ordering of $n$ for charmonia states, $n(\jpsi)>n(\psits)$, may be present, although the effect is smaller than for bottomonia.

Based on these observations, the values of $n$ for \uds and ground-state \qum mesons measured at midrapidity are fit. The selected data sets are indicated with circles around their markers in Figure~\ref{fig:individual_fits}. There are 12, 5, and 3 data samples at $\sqs=7$, 8, and 13~TeV, respectively. The $\sqs=7$ TeV values are fit to a linear function which becomes a  constant for a chosen value of parameter $T$.  Due to the low number of selected data sets at higher energies, the 8 and 13 TeV data are fit to a constant. The values of $n\left(\sqs=7,8,13\mathrm{[TeV]} \right) = \left(6.65, 6.34, 5.44\right)$, and are used in the analysis described below. These values are shown in Figure~\ref{fig:individual_fits} by horizontal lines. The quality of the fits and statistical uncertainties associated with the parameters are investigated in detail in a number of similar studies (e.g.~\cite{Grigoryan:2017gcg,Biro:2017eip}). In this analysis, the robustness of the common fits are checked by changing $T$ by $\pm50$ MeV, which shifts the points by approximately a unit but does not change the discussion or conclusions below.

\begin{figure}[htb]
\includegraphics[width=0.8\textwidth]{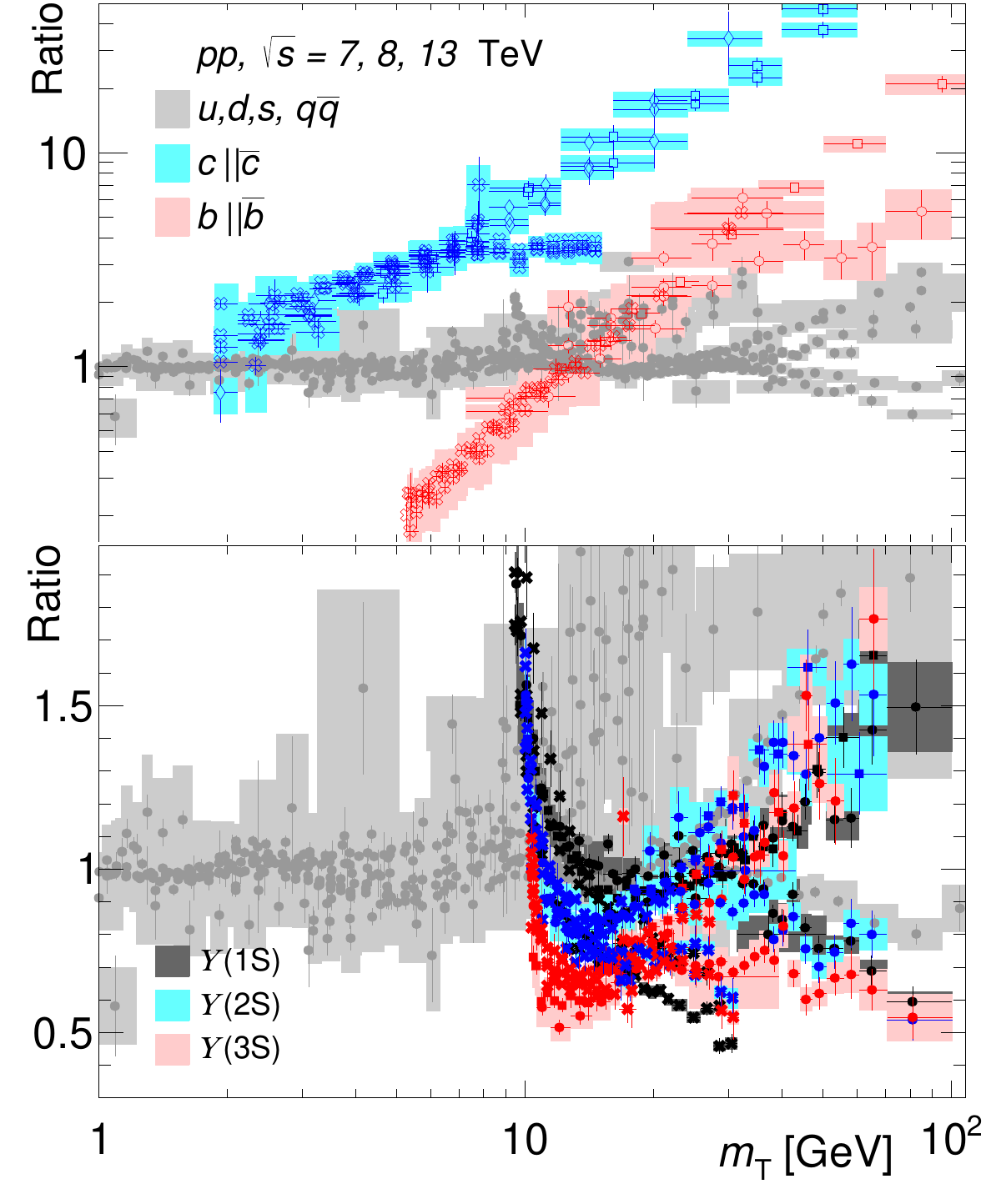}
\caption{\label{fig:common_fit} Measurements used in the common fit are shown in gray. Symbols follow the notation of Figure~\ref{fig:individual_fits}. Individual curves of \ohf are scaled with arbitrary factors for visibility.}
\end{figure}

Figure~\ref{fig:common_fit} shows the various measurements from all three collision energies, divided by the expectation from \mT-scaling in which $n$ at each collision energy is fixed to the values mentioned above.% fits to \uds and ground-state \qum meson. 
The points which are used in the common fit are shown in gray, other points are shown in colors. Overall, points shown in gray demonstrate reasonable agreement with unity, although at high-\pT \qum spectra tend to rise deviating by up to a factor of 2, somewhat similar to what can be seen in Ref.~\cite{Grigoryan:2017gcg}. Increasing $n(\sqs)$ can partially solve this problem, however, further optimizing the fit would be difficult without knowing the prompt components in the cross sections of heavier species. Taking this into account, and also given the ranges that the measurements cover in particle mass and \pT, the agreement to a common fit shown in the plot can be considered sufficient for the purposes of this analysis.

The upper panel of Figure~\ref{fig:common_fit} compares results measured for \ochm and \obtm with the results of the common fit. Individual \ohf measurements are vertically scaled with arbitrarily-chosen factors in the range 0.5--5 for the clarity of the plot. Differences in the slopes are expressions of the observation that \obtm has harder spectra, compared to \ochm and other particles, as seen in Figure~\ref{fig:individual_fits}. 

A comparison of the measured \btum and \ebtum states to the common fit is shown in the lower panel of Figure~\ref{fig:common_fit}. This figure shows all data, including measurements at forward rapidity and data points at $\pT<5$~GeV which are excluded from the fits. The splitting of \Upsn points at high-\mT is due to the $\sqs=13$~TeV measurements, whose $n$ is not accurately reflected in the common fit due to a lack of experimental data as seen in Figure~\ref{fig:individual_fits}. A significant rise for all \Upsn states at low \pT, i.e., \mT only slightly above the \Upsa mass, is clearly visible. The contribution of $\chi_{b}\mathrm{(1P)}\rightarrow\Uone$ has been estimated using \Pythia \cite{Sjostrand:2014zea} simulations %and these qualitatively may explain the sharp rise.
and found to be qualitatively consistent with the large excess at \mT close to the \Uone mass seen in the Figure. 
Higher $\chi_{b}$ states have not been simulated, but the similarity of the excess for \Utwo and \Uthree in the ratio at \mT close to the \Upsa mass suggests that a significant contribution also of these likely comes from $\chi_{b}$ decays and the non-prompt fraction of the different \Upsn states is similar as expected~\cite{Andronic:2015wma,LHCb:2014ngh}.
To check the effects of the non-prompt fraction varying between different \Upsn states, the \Pythia simulation for the $\chi_{b}\mathrm{(1P)}\rightarrow\Upsn$ processes has been modeled with the same rate as the non-prompt yield of \Uone. Then the non-prompt yields into \Uone were increased by an additional 20\%, i.e., nearly doubled. This is found not to affect the conclusions.

Since the meson ratios at high \pT are 
constant~\cite{ATLAS:2016kwu,ATLAS:2012lmu,CMS:2015lbl,CMS:2011rxs,CMS:2015xqv,LHCb:2014ngh}, \ebtum results shown in the figure are normalized to \btum at high-\pT by scaling factors. All \Utwo results are scaled by 0.9, and all \Uthree results by 0.7. The lower panel of Figure~\ref{fig:common_fit} shows that, apart from the rise at low \pT, \Uone follows the trend of particles included in the common fit. On the other hand, compared to \btum there is a clear deficit of \ebtum states at low and intermediate \pT.

\section{\label{sec:discussion}Discussion}
The \mT-scaling phenomenon observed at lower \sqs and at LHC energies for \uds mesons approximately holds for heavier \qum particles in their ground state. Analysis of the LHC data shows that mesons obey or deviate from the scaling depending on the heaviest quark contained in the particle and whether quarks forming the meson are bound in the ground or excited state. Particles with an open charm or bottom have significantly harder spectra than all other particles, and \obtm mesons have a harder spectra than \ochm mesons. 

\begin{figure}[htb]
\includegraphics[width=0.8\textwidth]{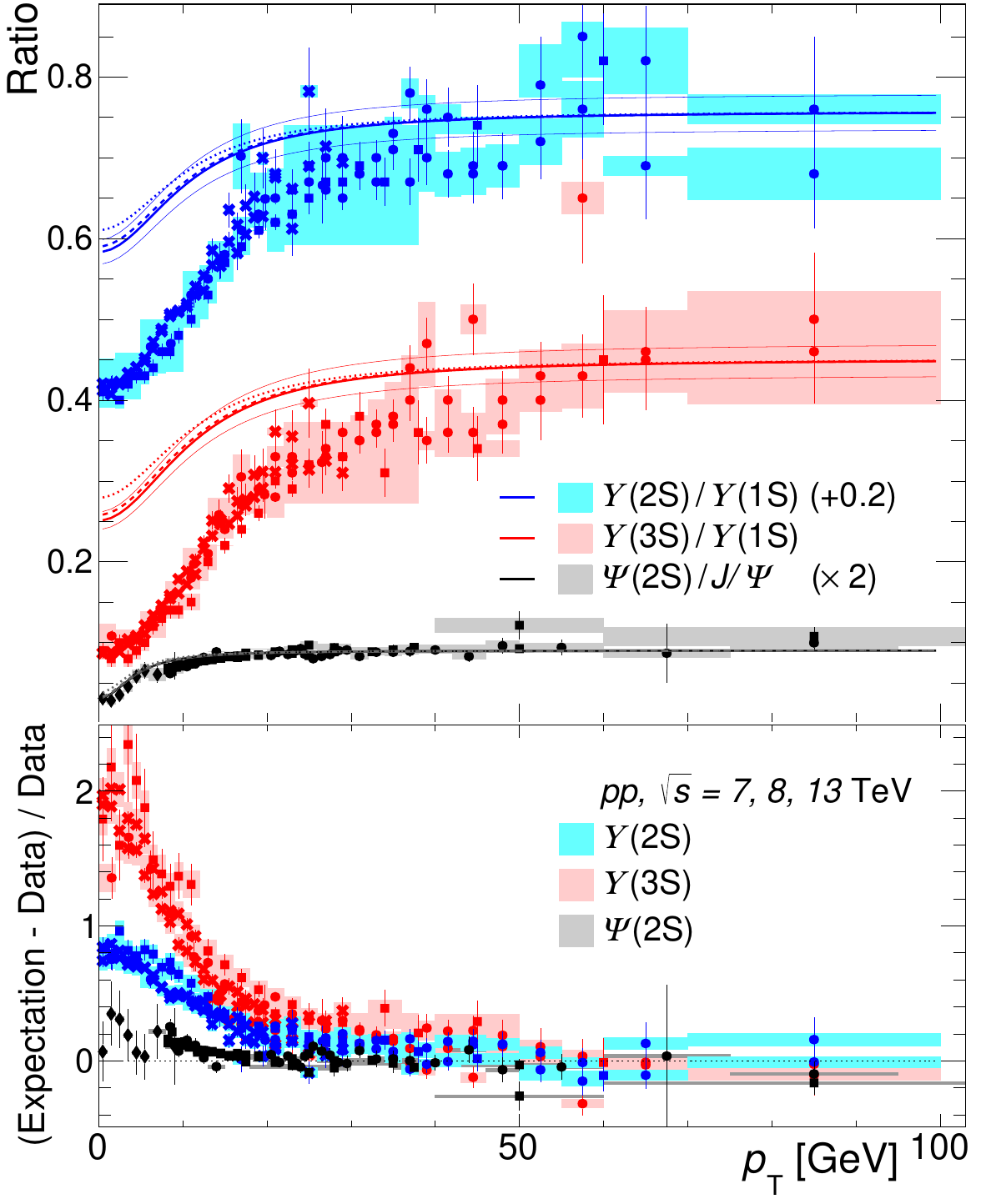}
\caption{\label{fig:ratios} Upper panel: Measured $\equm/\qum$ ratios are shown with markers. \psits data is scaled, and \Utwo data is shifted for plot clarity. Lines are expectations from the \mT-scaling prediction, normalized to the data in the region of $\pT>50$ GeV. The dashed (dotted) lines correspond $\sqs=8$ (13) TeV, and finer lines indicate the normalization uncertainty. Lower panel: The difference of the expected value based on \mT-scaling to the measurement divided by the measurement.}
\end{figure}

Employing \mT-scaling to compare particle spectra distributions is a commonly-used technique, as it reproduces well the momentum dependence of a variety of experimentally measured particles~\cite{ALICE:2020jsh}. In this letter, \mT-scaling is used to study the ratios of $\equm/\qum$ as is shown in the upper panel of Figure~\ref{fig:ratios}. 

Since the particle ratios are known to only weakly depend on rapidity and \sqs, all available LHC data is shown in the same plot. The \pT dependence of the ratios is not trivial and is challenging for models to reproduce~\cite{Andronic:2015wma}. In this letter, the data is compared to expectations based on \mT-scaling, which are shown with solid lines for the three collision energies. The \mT expectations are normalized to data at $\pT>50$~GeV, and the uncertainties resulting from the normalization are shown with thin lines. There are clear differences between the expectation and the data for \ebtum species and a significantly better consistency for \echum. 

To quantify the discrepancies, the lower panel shows the particles' `missing fraction' constructed from the ratios.  It is the measured value subtracted from the expected value and normalized to the measured, this normalization cancels the ground state production rate, and thus the quantity represents the `missing' excited state production equal to $\equm_{\mathrm{expected}} / \equm_{\mathrm{measured}} - 1$. The normalization uncertainty is added to the systematic uncertainty drawn at each point. The curves show that assuming the \mT-scaling scenario and \Uone-meson cross-section, at low \pT the \Utwo production expectation is approximately twice as high as the measurement and \Uthree is roughly three times higher. Estimation for \psits at low \pT is above zero, suggesting that the \echum may also be affected, although to a much lesser extent than the \Upsa case. With the existing data and understanding of the prompt fraction in \jpsi, a conclusion about \echum cannot be drawn with a large degree of certainty. At higher \pT, \Utwo and \Uthree points are significantly above zero to at least 30~GeV.

The observations reported here may share a common origin with the results reported in~\cite{CMS:2020fae}.
The CMS collaboration measured the production of \Upsn mesons at $\sqs=7$ GeV as a function of the charged particle multiplicity in the underlying event. CMS observed that events containing \Uone meson measure about two more tracks than events with \Uthree and one more track than events with \Utwo. The CMS collaboration noted that suppression of the \ebtum states at high multiplicity in \pp collisions can produce the measured effect. This assumption is qualitatively consistent with the discrepancy between the \ebtum production rates measured by the LHC experiments and their expectations from the \mT-scaling assumption that is plotted in Figure~\ref{fig:ratios}.  

If fewer tracks in collisions with \ebtum as observed by CMS, and reduced production of higher \Upsn mesons shown in Figure~\ref{fig:ratios} are two manifestations of the same physics process, it would mean that the measured cross-sections of \Utwo and \Uthree mesons are `suppressed' by factors of 1.6 and 2.4 compared to the expectation from the production rates of \Uone scaled by the missing fraction shown in the lower panel of Figure~\ref{fig:ratios}. A 5\% uncertainty on these estimates is associated with the normalization uncertainties, and other effects, which are expressed as deviations from the common fit as discussed above, may have a comparable impact on the values. Nevertheless, they are too small to explain the gap between the \mT-scaling expected and measured ratios shown in the upper panel of Figure~\ref{fig:ratios}.  The mechanism that suppresses the rates of \ebtum may also affect production rates of other \equm states, for example, \psits shown in Figure~\ref{fig:ratios}, however, in this particular case the effect is visibly weaker. Any conclusion about modification of the production rate of the \Uone-meson itself, lies outside of the framework of this analysis, because it breaks the \mT-scaling assumption. Existing theoretical cross-section calculations~\cite{Andronic:2015wma,Han:2014kxa,Abdulov:2020nxh,Abdulov:2020edt,Abdulov:2019uyx} show that the calculations generally better agree with the data for \Uone, compared to higher \Upsn, but these calculations are not available in the low-\pT region, where the discrepancy is the largest. Recently the comover interaction model successfully reproduced~\cite{Esposito:2020ywk} event multiplicity dependence of the \Upsn ratios measured by CMS. Theoretical interpretations of the effect that would simultaneously reproduce the suppression of excited \Upsn states and reduction of tracks in the collisions are needed.

\section{\label{sec:summary}Summary}
A systematic analysis of transverse-momentum distributions of mesons produced in \pp collisions at LHC energies suggests that the spectra of light mesons and ground state quarkonia particles have the same spectral shape, given by Eq.~\eqref{eq:mT} with universal parameters $T$ and $n$ for a given collision energy. Open-heavy-flavor mesons with $c$-quark in them have significantly harder spectra, and open-flavor $b$-mesons even harder. Production rates of excited bottomonia states with respect to ground state are significantly suppressed at \pT approaching zero, with more suppression measured for the higher excited state. The effect diminishes with increasing \pT but remains significant at least up to 30 GeV. Production rates of \Utwo and \Uthree states are fewer by factors 1.6 and 2.4, respectively, than they would be anticipated from the \mT-scaling assumption. This suppression of higher \Upsn  qualitatively agrees with the effect observed in~\cite{CMS:2013jsu,CMS:2020fae} and may be a manifestation of the same physics phenomenon. Production rate of \psits is generally consistent with the expectation from \jpsi, although a smaller-magnitude suppression cannot be excluded. Current theoretical calculations  can be advanced by relating the cross-section measurements to the event multiplicity. 

\begin{acknowledgments}
Research of Z.C. is supported by the Israel Science Foundation (grant 1946/18). Research of I.A. and A.M. is supported by Israel Academy of Science and Humanity and the MINERVA Stiftung with the funds from the BMBF of the Federal Republic of Germany.
\end{acknowledgments}

\bibliography{mtscaling}

\end{document}